\documentclass[fleqn,usenatbib,useAMS]{mnras}

\usepackage{graphicx}	% Including figure files
\usepackage{amsmath}	% Advanced maths commands
\usepackage{amssymb}	% Extra maths symbols
\usepackage{multicol}        % Multi-column entries in tables
\usepackage{bm}		% Bold maths symbols, including upright Greek
\usepackage{pdflscape}	% Landscape pages

\usepackage[T1]{fontenc}
\usepackage{ae,aecompl}

\usepackage{newtxtext,newtxmath}
\title{Ephemeris Updates for Seven Selected HATNet Survey Transiting Exoplanets}

\author[Poro et al.]{A. Poro$^{1,}$\thanks{Corresponding author: poroatila@gmail.com}%
, F. Ahangarani Farahani$^{2}$,
E. Jahangiri$^{2}$,
A. Sarostad$^{2}$,
M. Gozarandi$^{2}$,
M. Haghgou$^{2,3}$,\newauthor
F. Abolhassani$^{2}$,
A. Fakhrabadi$^{2}$,
Y. Jongen$^{4}$,
A. Wünsche$^{5}$,
R. Naves$^{6}$,
P. Guerra$^{7}$,
A. Marchini$^{8}$,\newauthor
M. Salisbury$^{9}$,
R. Ehrenberger$^{10}$,
V-P. Hentunen$^{11}$
\\
\\
$^{1}$Astronomy Department of the Raderon AI Lab., BC., Burnaby, Canada\\
$^{2}$BSN Project, Transiting Exoplanets Department, Tehran, Iran\\
$^{3}$Department of Chemistry, Faculty of Science, University of Guilan, Rasht, Iran\\
$^{4}$Observatoire de Vaison-La-Romaine, Departementale 51, 84110 Vaison-La-Romaine, France\\
$^{5}$Observatoire des Baronnies Provençales, Route de Nyons, F-05150 Moydans, France\\
$^{6}$Observatorio Montcabrer, Cabrils, Spain\\
$^{7}$Observatori Astronòmic Albanyà, Camí de Bassegoda S/N, Albanyà E-17733, Girona, Spain\\
$^{8}$University of Siena - Astronomical Observatory, Via Roma 56, 53100 Siena, Italy\\
$^{9}$British Astronomical Association, 25 Farringdon Street, London EC4A 4AB, England\\
$^{10}$Czech Astronomical Society, Fričova 298, 251 65 Ondřejov, Czech Republic\\
$^{11}$Taurus Hill Observatory, 79480 Varkaus, Finland\\}

\date{Accepted ---. Received ---; in original form ---}

\pubyear{2023}

\begin{document}
\label{firstpage}
\pagerange{\pageref{firstpage}--\pageref{lastpage}}
\maketitle

\begin{abstract}
We refined the ephemeris of seven transiting exoplanets HAT-P-6b, HAT-P-12b, HAT-P-18b, HAT-P-22b, HAT-P-32b, HAT-P-33b, and HAT-P-52b. We observed 11 transits from eight observatories in different filters for HAT-P-6b and HAT-P-32b. Also, the Exoplanet Transit Database (ETD) observations for each of the seven exoplanets were analyzed, and the light curves of five systems were studied using Transiting light Exoplanet Survey Satellite (TESS) data. We used Exofast-v1 to simulate these ground- and space-based light curves and estimate mid-transit times. We obtained a total of 11, 175 and 67 mid-transit times for these seven exoplanets from our observations, ETD and TESS data, respectively, along with 155 mid-transit times from the literature. Then, we generated transit timing variation (TTV) diagrams for each using derived mid-transit times as well as those found in the literature. The systems' linear ephemeris was then refined and improved using the Markov Chain Monte Carlo (MCMC) method. All of the studied exoplanets, with the exception of the HAT-P-12b system, displayed an increasing trend in the orbital period in the TTV diagrams.
\end{abstract}


%%%%%%%%%%%%%%%%%%%%%%%%%%%%%%%%%%%%%%%%%%%%%%%%%%

\section{Introduction}
The number of exoplanets discovered and characterized each year has been increasing since the results of the first exoplanet detection [1]. Hot Jupiters are an important type of planetary gas giant with masses and radii similar to Jupiter but orbiting their host stars with short orbital periods (most less than 10 days), making them a good target system to discover and study [2][3].
The transit technique is the most efficient way to improve our understanding of exoplanets through ground- and space-based surveys. Furthermore, photometric transit surveys combined with radial velocity data have become one of the most successful methods for detecting transiting exoplanets over the past decade [4]. High-precision transit observations provide information to refine planetary parameters such as the planet’s size, mass, atmosphere, and orbital ephemerides [5][6]. Moreover, photometric transit surveys allow us to study the variations of the orbital periods through TTV analysis.
Space telescopes have longer available observational time, and they are not affected by the Earth’s atmosphere as well. TESS is one of the most significant space-based survey missions for the discovery and observation of transiting exoplanets. TESS was launched in 2018 to observe new exoplanets orbiting bright nearby stars that are brighter than Kepler mission stars [7]. Furthermore, when combined with previous work, this space mission provides precise transit timing for discovered exoplanets, which is critical for obtaining a better transit ephemeris [8].

Based on our observations, TESS, ETD\footnote{\url{http://var2.astro.cz/ETD/}}, and literature observations, we updated orbital ephemeris for the HAT-P-6b, HAT-P-12b (TESS ID 198108326), HAT-P-18b (TESS ID 21744120), HAT-P-22b (TESS ID 252479260), HAT-P-32b, HAT-P-33b (TESS ID 239154970), and HAT-P-52b (TESS ID 436875934). These exoplanets were discovered by the Hungarian-made Automated Telescope Network (HATNet) survey.

%--------------------------------------------------------------------

\section{Observations and method}
\subsection{Observation and data reduction}
Observations in this study have been made regarding exoplanets HAT-P-6b and HAT-P-32b during the years 2018 to 2022. A total of nine observation nights have been done for these two exoplanets; five and four nights for HAT-P-6b and HAT-P-32b, respectively. All these photometric observations have been done with small telescopes at eight observatories. We used CCD and standard filters in these observations. The information about the observatories, telescopes, CCDs, and data reduction software that we used is listed in Table \ref{tab1}. In Table \ref{tab1}, an abbreviated name has been determined for each observatory just to identify them in this study.
The basic data reduction for the dark, bias, and flat field of each CCD image was carried out in accordance with the standard technique.

\subsection{ETD data}
To obtain the refined orbital ephemeris of selected HATNet exoplanets, we also collected also light curves, which were sourced from astronomers through the ETD archive [9]. Light curves were obtained from various filters and time scales. We used data in ETD that we were confident enough to be appropriate; for example, we did not use data whose declared time was less than three digits. We used those which generally have a quality index (DQ) of less than three [9]. All times in the data were converted from JD or HJD to $BJD_{TDB}$ based on the geographic location of observation and RA(J2000) and DEC(J2000) from the Simbad\footnote{\url{http://simbad.u-strasbg.fr/simbad/}}  astronomical database.

In some ETD light curves, the airmass effect has been ignored, so airmass must be calculated based on the observers’ location, which influences and improves the measured mid-transit times of related light curves. Therefore, we computed the airmass using the Astropy package in Python [10].

\subsection{TESS data}
Five of these exoplanets were observed by TESS, and HAT-P-6b, and HAT-P-32b have no TESS data yet. TESS observed the five host stars at 120-second cadences. We collected TESS data from the Mikulski space telescope archive (MAST). TESS style curves were extracted by LightKurve\footnote{\url{https://docs.lightkurve.org/}} code  from the MAST Python package.

\subsection{Method}
\begin{figure*}
\begin{center}
   \includegraphics[scale=0.44]{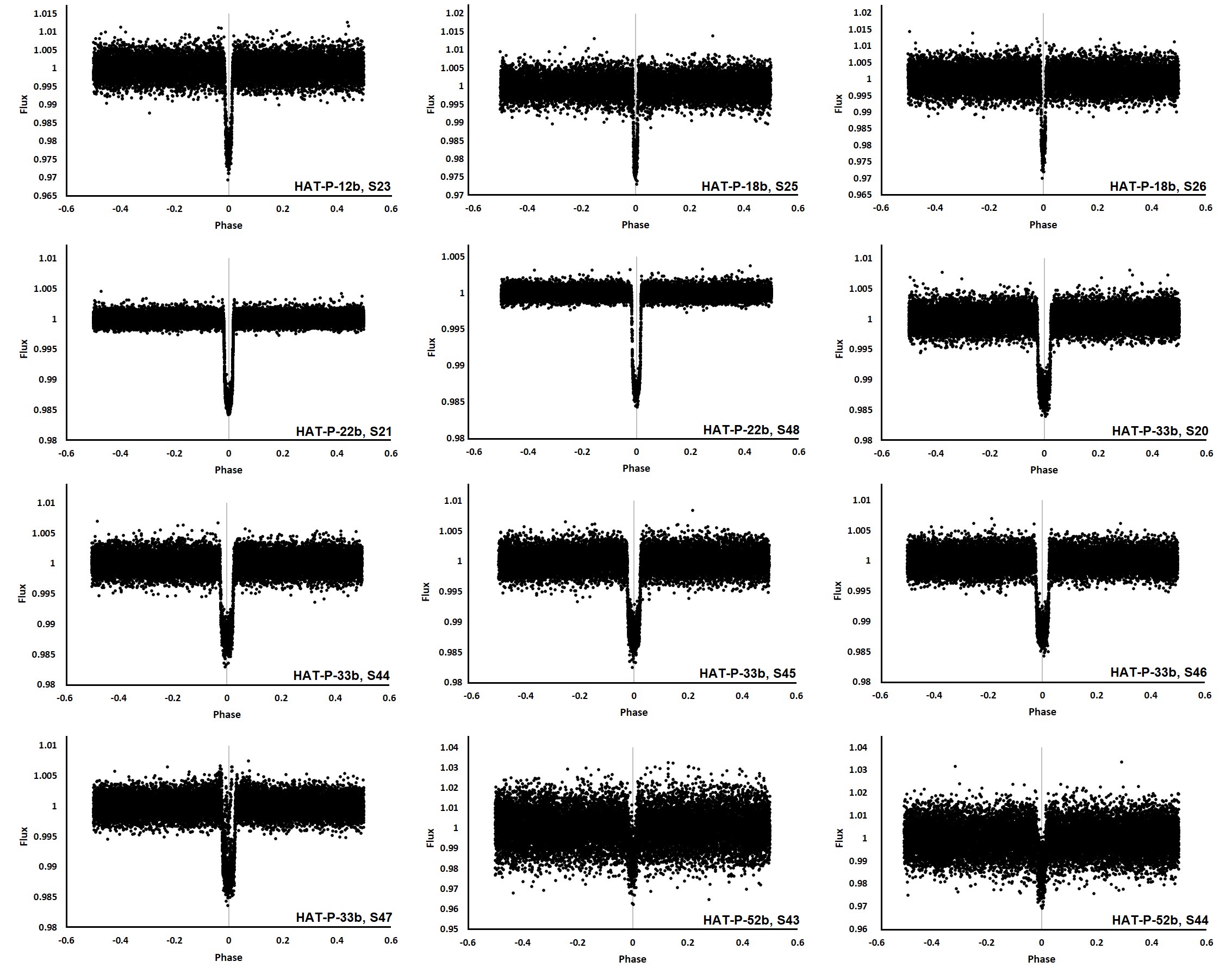}
   \caption{Folded TESS light curves in each sector of all selected exoplanets were obtained from the LightKurve code.}
\label{Fig1}
\end{center}
\end{figure*}

\begin{figure*}
\begin{center}
   \includegraphics[scale=0.30]{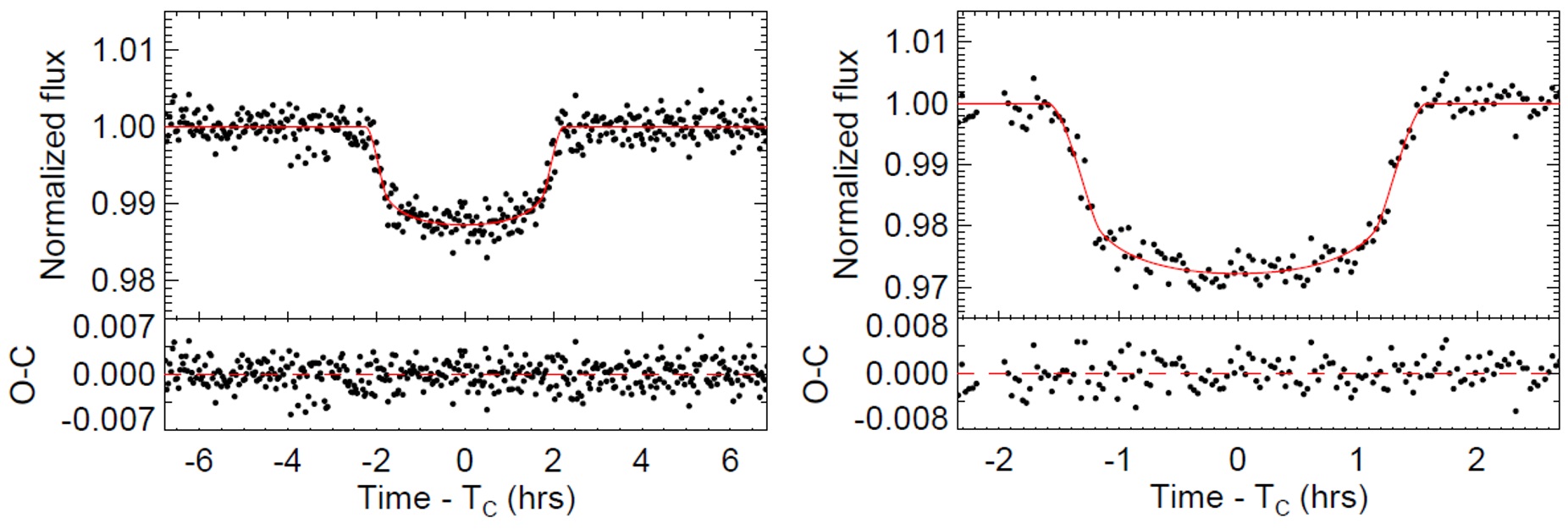}
   \caption{Left: HAT-P-33b observational and theoretical light curves using TESS sector 45 data; Right: The observational light curve of HAT-P-32b from this study in the $V$ filter and the theoretical light curve.}
\label{Fig2}
\end{center}
\end{figure*}

\begin{figure*}
\begin{center}
\includegraphics[scale=0.44]{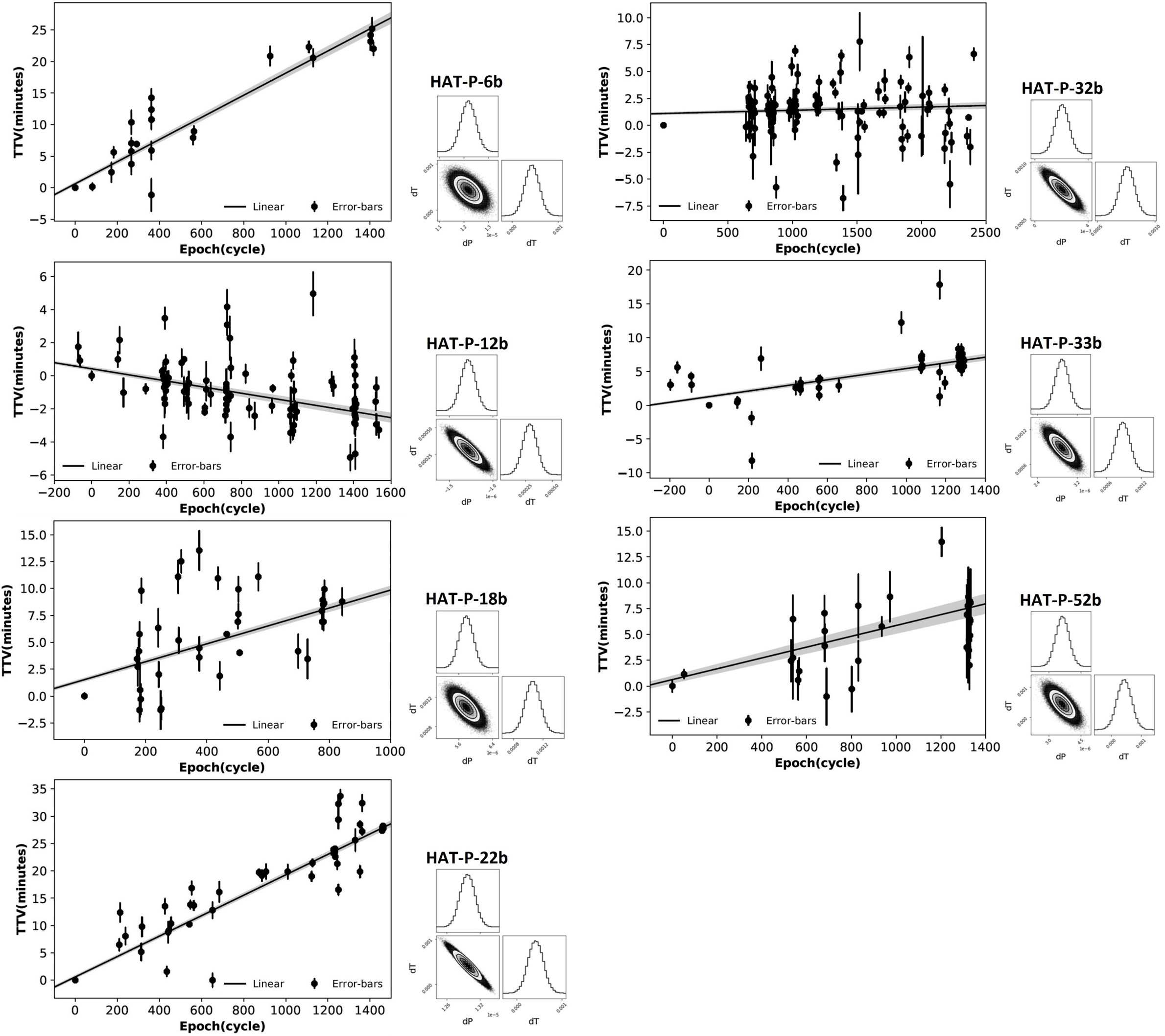}
    \caption{The TTV diagrams of seven studied exoplanets with the linear fit on the data points and posterior distributions for the fitted parameters using MCMC (dT and dP).}
\label{Fig3}
\end{center}
\end{figure*}

We relied on the AstroImageJ software [11] to normalize all of the data. Figure \ref{Fig1} shows the folded TESS light curves for five selected exoplanets. Finally, all ground- and space-based light curves were applied to Exofast-v1 [12] for modeling purposes; as a consequence, the output mid-transit times and associated uncertainties were employed. Figure \ref{Fig2} provides an example of a modeled observation of TESS and this study's observation. The extracted transit mid-times from our observations and TESS data are provided in Tables \ref{tab2} and \ref{tab3}. Tables \ref{tab5}-\ref{tab11} include the literature and ETD transit mid-times.
\\
\\
We plotted TTV diagrams for seven selected exoplanets using derived mid-transit times and those available within the literature. Our MCMC analysis of these timings enabled us to refine the linear ephemeris of the systems. We applied the MCMC method, or sampling from the posterior probability distributions of the coefficients (100 walkers, 10000 step number, and 1000 burn-in) using the Pymc3 package in Python [13].  Figure \ref{Fig3} shows all TTV-diagrams of studied exoplanets and also displays the posterior distributions for the fitted parameters using the MCMC method (dT and dP).

\begin{table*}
\caption{The observatories of this study and the instruments that were employed.}
\centering
\begin{center}
\footnotesize
\begin{tabular}{c c c c}
 \hline
 \hline
 Observatory		& Telescope		& CCD		& Data reduction Software\\
\hline
Rasteau Observatory, France	(RO) &	PlaneWave CDK 17"	&	SBIG STXL11004	&	Muniwin / C-munipack	\\
Montcabrer private observatory, Spain	(MO) &	SCT 12"	&	SBIG ST8-XME	&	Fotodif	\\
Observatori Astronomic Albanyà, Spain	(AA) &	Meade ACF 16"	&	Moravian Instruments G4-9000	&	Fotodif	\\
Astronomical Observatory, University of Siena (K54), Italy	(AO) &	MCT 300 mm	&	SBIG STL-6303	&	Muniwin / C-munipack \\
Observatoire des Baronnies Provençales, France	(BO) &	Cassegrain 430 mm	&	Zwo ASI6200 Pro mono	&	Muniwin / C-munipack	\\
Private Observatory, Czech Republic	(PO) &	400 mm	&	SBIG ST-10 XME	&	AstroImageJ 3.2.10	\\
Crow-Observatory Vranová, Czech Republic	(CO) &	NWT 300 mm	&	Moravian Instruments G2-3200	&	Muniwin / C-munipack	\\
Taurus Hill Observatory, Finland	(TO) &	SCT 14"	&	SBIG ST-8 XME	&	AIP4Win v2.4.10	\\
\hline
\hline
\end{tabular}
\end{center}
\label{tab1}
\end{table*}

\begin{table*}
\caption{Extracted ground-based transit times for HAT-P-6b and HAT-P-32b in this study.}
\centering
\begin{center}
\footnotesize
\begin{tabular}{c c c c c c c}
 \hline
 \hline
 Exoplanet	& Observatory	& $T_c$($BJD_{TDB}$)	& Error	& Filter	& Eopch	& O-C\\
\hline
HAT-P-6b	&	MO	&	2455430.46458	&	0.00110	&	Optec $R$ cousins	&	362	&	0.0075	\\
HAT-P-6b	&	RO	&	2458312.50536	&	0.00063	&	Baader imaging $G$	&	1110	&	0.0155	\\
HAT-P-6b	&	AA	&	2458389.56389	&	0.00101	&	Baader $J-C V$	&	1130	&	0.0143	\\
HAT-P-6b	&	AO	&	2459441.43053	&	0.00101	&	Johnsons-Cousins $I$	&	1403	&	0.0161	\\
HAT-P-6b	&	RO	&	2459441.43128	&	0.00095	&	Baader imaging $R$	&	1403	&	0.0168	\\
HAT-P-6b	&	RO	&	2459468.40283	&	0.00124	&	Baader imaging $R$	&	1410	&	0.0175	\\
HAT-P-32b	&	CO	&	2459107.46235	&	0.00097	&	Johnsons-Cousins $R_c$	&	2180	&	-0.0015	\\
HAT-P-32b	&	TO	&	2459191.31427	&	0.00034	&	Baader Bessell photometric $R$	&	2219	&	0.0001	\\
HAT-P-32b	&	BO	&	2459507.36578	&	0.00017	&	Johnsons-Cousins $V$	&	2366	&	0.0005	\\
HAT-P-32b	&	PO	&	2459593.36707	&	0.00024	&	Johnsons-Cousins $R_c$	&	2406	&	0.0015	\\
HAT-P-32b	&	BO	&	2459593.37024	&	0.00041	&	Johnsons-Cousins $R_c$	&	2406	&	0.0046	\\
\hline
\hline
\end{tabular}
\end{center}
\label{tab2}
\end{table*}

\begin{table*}
\caption{Extracted TESS transit times for five exoplanets.}
\centering
\begin{center}
\footnotesize
\begin{tabular}{c c c c c c c c c c}
 \hline
 \hline
 Exoplanet & $T_c(BJD_{TDB})$ & Error & Epoch & O-C &  Exoplanet & $T_c(BJD_{TDB})$ & Error & Epoch & O-C\\
\hline
HAT-P-12b	&	2458933.54320	&	0.00053	&	1405	&	-0.0014	&	HAT-P-33b	&	2459506.13961	&	0.00051	&	1265	&	0.0040	\\
HAT-P-12b	&	2458936.75601	&	0.00052	&	1406	&	-0.0016	&	HAT-P-33b	&	2459509.61506	&	0.00052	&	1266	&	0.0050	\\
HAT-P-12b	&	2458939.96969	&	0.00048	&	1407	&	-0.0010	&	HAT-P-33b	&	2459516.56424	&	0.00058	&	1268	&	0.0053	\\
HAT-P-12b	&	2458946.39512	&	0.00055	&	1409	&	-0.0017	&	HAT-P-33b	&	2459520.03824	&	0.00054	&	1269	&	0.0048	\\
HAT-P-12b	&	2458949.60784	&	0.00050	&	1410	&	-0.0020	&	HAT-P-33b	&	2459523.51230	&	0.00048	&	1270	&	0.0044	\\
HAT-P-12b	&	2458952.82114	&	0.00048	&	1411	&	-0.0018	&	HAT-P-33b	&	2459526.98761	&	0.00052	&	1271	&	0.0052	\\
HAT-P-18b	&	2458989.25305	&	0.00059	&	776	&	0.0055	&	HAT-P-33b	&	2459530.46206	&	0.00057	&	1272	&	0.0052	\\
HAT-P-18b	&	2458994.76038	&	0.00048	&	777	&	0.0048	&	HAT-P-33b	&	2459533.93584	&	0.00055	&	1273	&	0.0045	\\
HAT-P-18b	&	2459005.77787	&	0.00050	&	779	&	0.0062	&	HAT-P-33b	&	2459537.41024	&	0.00057	&	1274	&	0.0044	\\
HAT-P-18b	&	2459011.28554	&	0.00051	&	780	&	0.0059	&	HAT-P-33b	&	2459540.88529	&	0.00065	&	1275	&	0.0050	\\
HAT-P-18b	&	2459016.79255	&	0.00059	&	781	&	0.0048	&	HAT-P-33b	&	2459544.35953	&	0.00062	&	1276	&	0.0048	\\
HAT-P-18b	&	2459027.80975	&	0.00060	&	783	&	0.0060	&	HAT-P-33b	&	2459547.83335	&	0.00063	&	1277	&	0.0041	\\
HAT-P-18b	&	2459033.31868	&	0.00060	&	784	&	0.0069	&	HAT-P-33b	&	2459554.78399	&	0.00053	&	1279	&	0.0058	\\
HAT-P-22b	&	2458871.63055	&	0.00020	&	1227	&	0.0166	&	HAT-P-33b	&	2459558.25712	&	0.00055	&	1280	&	0.0044	\\
HAT-P-22b	&	2458874.84229	&	0.00018	&	1228	&	0.0161	&	HAT-P-33b	&	2459561.73071	&	0.00057	&	1281	&	0.0036	\\
HAT-P-22b	&	2458878.05478	&	0.00018	&	1229	&	0.0164	&	HAT-P-33b	&	2459568.68054	&	0.00053	&	1283	&	0.0044	\\
HAT-P-22b	&	2458881.26687	&	0.00019	&	1230	&	0.0163	&	HAT-P-33b	&	2459572.15466	&	0.00054	&	1284	&	0.0041	\\
HAT-P-22b	&	2458887.69120	&	0.00017	&	1232	&	0.0161	&	HAT-P-33b	&	2459575.63036	&	0.00053	&	1285	&	0.0053	\\
HAT-P-22b	&	2458890.90361	&	0.00020	&	1233	&	0.0163	&	HAT-P-33b	&	2459582.57829	&	0.00053	&	1287	&	0.0043	\\
HAT-P-22b	&	2458894.11620	&	0.00018	&	1234	&	0.0167	&	HAT-P-33b	&	2459586.05330	&	0.00053	&	1288	&	0.0048	\\
HAT-P-22b	&	2458897.32739	&	0.00020	&	1235	&	0.0157	&	HAT-P-33b	&	2459589.52750	&	0.00051	&	1289	&	0.0046	\\
HAT-P-22b	&	2459613.65575	&	0.00019	&	1458	&	0.0190	&	HAT-P-33b	&	2459599.95038	&	0.00055	&	1292	&	0.0040	\\
HAT-P-22b	&	2459616.86844	&	0.00018	&	1459	&	0.0194	&	HAT-P-33b	&	2459603.42543	&	0.00053	&	1293	&	0.0046	\\
HAT-P-22b	&	2459620.08042	&	0.00018	&	1460	&	0.0192	&	HAT-P-52b	&	2459475.83726	&	0.00205	&	1316	&	0.0026	\\
HAT-P-22b	&	2459626.50524	&	0.00016	&	1462	&	0.0196	&	HAT-P-52b	&	2459478.59303	&	0.00204	&	1317	&	0.0048	\\
HAT-P-22b	&	2459629.71736	&	0.00016	&	1463	&	0.0195	&	HAT-P-52b	&	2459489.60804	&	0.00179	&	1321	&	0.0054	\\
HAT-P-22b	&	2459632.92936	&	0.00017	&	1464	&	0.0193	&	HAT-P-52b	&	2459492.35880	&	0.00239	&	1322	&	0.0026	\\
HAT-P-33b	&	2458845.98935	&	0.00055	&	1075	&	0.0038	&	HAT-P-52b	&	2459495.11585	&	0.00202	&	1323	&	0.0060	\\
HAT-P-33b	&	2458849.46501	&	0.00051	&	1076	&	0.0050	&	HAT-P-52b	&	2459500.61943	&	0.00171	&	1325	&	0.0024	\\
HAT-P-33b	&	2458852.93919	&	0.00053	&	1077	&	0.0047	&	HAT-P-52b	&	2459503.37368	&	0.00163	&	1326	&	0.0031	\\
HAT-P-33b	&	2458859.88850	&	0.00051	&	1079	&	0.0051	&	HAT-P-52b	&	2459506.12841	&	0.00149	&	1327	&	0.0042	\\
HAT-P-33b	&	2458863.36179	&	0.00051	&	1080	&	0.0039	&	HAT-P-52b	&	2459508.87918	&	0.00163	&	1328	&	0.0014	\\
HAT-P-33b	&	2458866.83641	&	0.00054	&	1081	&	0.0041	&	HAT-P-52b	&	2459514.38955	&	0.00156	&	1330	&	0.0045	\\
HAT-P-33b	&	2459502.66693	&	0.00052	&	1264	&	0.0058	&		&		&		&		&		\\
\hline
\hline
\end{tabular}
\end{center}
\label{tab3}
\end{table*}

\section{Refined transit ephemeris}
Follow-up observations of known transiting exoplanets, either with photometry or high-resolution spectroscopy, are an important issue. Accordingly, a growing quantity of this data provides the refined physical planetary properties, formation, and evolution processes of these exoplanets. Since the orbital ephemeris of exoplanets demonstrates their places in their orbits, refinement of the exoplanet’s ephemeris would be a key factor for the prediction of future transit events.

Based on the reference ephemeris for each exoplanet, epochs, and O-Cs are calculated. According to extracted mid-transit times, we can compute a new ephemeris by a well-known linear relation (1),

\begin{equation}
\label{eq1}\begin{aligned}
T_c =T_0+E\times P
\end{aligned}
\end{equation}

where $T_0$ is considered as the reference mid-transit time, $P$ and $E$ are the orbital period and the number of epochs passed since $T_0$, respectively. New ephemeris and reference ephemeris for exoplanets are given in Table \ref{tab4}.

\begin{table*}
\caption{The new ephemeris derived by a linear fit on the TTV diagram of each exoplanet and reference ephemeris for computing epochs and the TTV values.}
\centering
\begin{center}
\footnotesize
\begin{tabular}{c c c c c}
 \hline
 \hline
 Exoplanet && New ephemeris ($BJD_{TDB})$ && Reference ephemeris ($BJD_{TDB}$)\\
\hline
HAT-P-6b &&
 $2454035.6769526(3)+3.85300(15)\times E$ && $2454035.67652(28)+3.852985(5)\times E$  [14]\\
\hline
HAT-P-12b &&
$2454419.19585(6)+3.21305852(8)\times E$ && $2454419.19556(20)+3.2130598(21)\times E$ [15]\\
\hline
HAT-P-18b &&
$2454715.022802(97)+5.5080288(2)\times E$ && $2454715.02174(20)+5.508023(6)\times E$ [16]\\
\hline
HAT-P-22b &&
$2454930.22043(16)+3.2122330(1)\times E$ && $2454930.22001(25)+3.212220(9)\times E$ [17]\\
\hline
HAT-P-32b &&
$2454420.44713(6)+2.15000821(5)\times E$ && $2454420.44637(9)+2.150008(1)\times E$ [16]\\
\hline
HAT-P-33b &&
$2455110.92683(12)+3.4744769(1)\times E$ && $2455110.92595(22)+3.474474(1)\times E$ [16]\\
\hline
HAT-P-52b &&
$2455852.10370(23)+2.7535989(3)\times E$ && $2455852.10326(41)+2.7535953(94)\times E$ [18]\\
\hline
\hline
\end{tabular}
\end{center}
\label{tab4}
\end{table*}

\subsection{HAT-P-6b}
The exoplanet HAT-P-6b is a hot Jupiter with a radius of 1.33±0.06 $R_J$ and a mass of 1.06±0.12 $M_J$ [14]. The [14] study, which was based on the primary transit method's results, announced its discovery. According to a report by [14] on the characteristics of the planet and its host star, e=0.046±0.031 indicates that the planet's orbit is almost circular. [19] collected transit light curves for this exoplanet and refitted available photometry data with their new and literature mid-transit times. A revised orbital period and mid-transit time for this exoplanet were also updated by [20].

We collected the high-quality mid-transit times that were previously published in the literature, and the mid-transit times resulted from the modeling of ETD light curves for plotting an updated TTV diagram. We extracted a total of six mid-transit times from our observations and 13 mid-transit from ETD. We note that this exoplanet has not had TESS data.

\subsection{HAT-P-12b}
[15] used the transit method to discover the exoplanet HAT-P-12b, which orbits a dwarf star According to the results of the discovery paper, HAT-P-12b is one of the least massive gas giant planets that was found until its discovery and is an exoplanet of the sub-Saturn type. The $J$ band photometry study by [21] of this exoplanet obtained the precise mid-transit time to constrain the transit-time variations of the HAT-P-12b system. [22] observed three new transit light curves of HAT-P-12b, and with existing literature data, they came up with an improved ephemeris. [22] also refined the absolute physical properties of the star-planet system. [23] updated the ephemeris of HAT-P-12b according to six transits for this system by applying a least-squares linear fit to all available transit times. According to their results, no long-term TTVs were apparent. [24] studied HAT-P-12b in bands $V$ and $I$ to investigate the transmission spectrum of this system. [24] observed 23 new photometric transit light curves, and analysis showed no indication of star-spot influence on the calculated transit parameters. [25] studied this exoplanet's atmosphere. In fact, the goal of this research was to specify an appropriate solution for future studies of other exoplanetary atmospheres. Spectroscopic observations using the Large Binocular Telescope (LBT) were done by [26] to obtain an atmosphere transmission spectrum of this exoplanet. They found no evidence of Na or K absorption features in the relatively flat transmission spectrum, which is in agreement with the HST transmission spectrum. Furthermore, [27] included six new mid-transit times to determine a new ephemeris by a linear fit to a satisfactory level. [28] also reported an infrared transmission photometry of HAT-P-12b with the other 48 exoplanets with the largest analysis of Spitzer/IRAC observations to study the influence of infrared photometry on atmospheric chemical properties.

We used mid-transit times conducted from the modeling of ETD light curves and TESS in association with data published in previous literature for plotting a new TTV diagram. We extracted 27 mid-transit times from ETD and 6 mid-transit times from sector 23 of TESS for HAT-P-12b.

\subsection{HAT-P-18b}
HAT-P-18b is a low-density Saturn-mass exoplanet orbiting a super-solar metallicity K2 dwarf star [29]. The discovery observations of this exoplanet have been made by [29] using the transit method to obtain the orbital and physical properties of the system. [29] reported a non-zero ($e=0.084±0.048$) eccentricity for HAT-P-18b. Complementary new photometric observations of the full transit were also analyzed by [30] in order to independently estimate the parameters of the host star and HAT-P-18b. [31] performed the TTV study of HAT-P-18b with a limited number of existing high-quality data and they presented ground-based transmission spectroscopy of HAT-P-18b. This exoplanet was described as a hot Jupiter by [32], who also found Rayleigh-scattering in the atmosphere, and their results confirmed that ground-based observations are suitable to determine the opacity sources of exoplanets’ atmospheres. [33] observed HAT-P-18b as a part of the original research by the young twinkle students (ORBYTS) program to refine its transit ephemerides. The atmosphere of this exoplanet has been studied by [34]. Moreover, [35] derived the refined ephemeris from observations provided by the ExoClock network in combination with previous literature data.

For HAT-P-18b, we obtained seven mid-transit times from sectors 25 and 26 of TESS, and 21 mid-transit times from ETD.

\subsection{HAT-P-22b}
[17] reported the discovery of the exoplanet HAT-P-22b. It is among the moderately massive and compact hot Jupiters, orbiting a fairly metal-rich dwarf star with a $V$=9.732 magnitude. [36] presented the first photometric follow-up observation of bright transiting exoplanets by using a defocusing technique. Following this, [37] performed a follow-up transit observation using a defocusing technique and they derived one complete transit and computed the mid-transit times for HAT-P-22b. The near-$UV$ and optical photometric observations of HAT-P-22b were made by [38] to study the atmosphere of this exoplanet. [38] also refined the planetary parameters and ephemerides of HAT-P-22b hot Jupiter. Accordingly, all derived parameters were in agreement with the discovery values by [17], and any non-spherical asymmetries were not seen in their data.

In order to plot a TTV diagram for this exoplanet, we extracted 30 and 14 mid-transit times resulting from the modeling of ETD and TESS light curves (sectors 21 and 48), respectively, as well as data from previous publications.

\subsection{HAT-P-32b}
The planet HAT-P-32b was discovered by the HATNet survey in 2011 and it is a hot Jupiter exoplanet orbiting a late-F-early-G dwarf star with $V$=11.289 magnitude. In this discovery, radial velocity measurements were taken with High-Resolution Echelle Spectrometer and [39] transit model was used in order to describe the HATNet photometry [16]. [21] presented a $JH$-band photometry observation of HAT-P-32b and extracted precise mid-transit times. [21] declared that HAT-P-32b system parameters were in agreement with those reported in the [16] study and derive a period of this exoplanet with improved uncertainty. Following this, [40] reported two primary transits of HAT-P-32b during Gemini-North Gemini Multi-Object Spectrograph observations. They used white light curve analysis in order to refine the parameters of this exoplanet and derive new ephemeris. [41] updated the system properties by analyzing the results of 45 transit observations, which were observed by using the young exoplanet transit initiative (YETI) network. Moreover, [41] studied the TTV diagram to investigate the existence of an additional planet in the HT-P-32b system. [42] performed a global fit for the HAT-P-32b system based on their new photometric observations and previously published RV data in order to update the system parameters. [42] also analyzed the TTV diagram for this system and according to the results, there was no significant TTV signals. Some follow-up high-quality observations of this exoplanet were done with small observatories operated by citizen scientists in 2020 [43].

The accurate mid-transit times for HAT-P-32b were obtained from the available data for plotting an updated TTV diagram. We extracted a total of five and 72 mid-transit times from our observations and the ETD, respectively.

\subsection{HAT-P-33b}
The planet HAT-P-33b was among the first exoplanets discovered by the HATNet survey in 2011 and was confirmed by high-precision photometry and additional radial velocity measurements [16]. HAT-P-33b is an inflated hot Jupiter orbiting a Late-F dwarf star with a short orbital period. [16] reported that HAT-P-33b has a radius of $\sim1.7$ $R_J$ which is among one of the largest measured radius for all transiting exoplanets. HAT-P-33b also has an equilibrium temperature of more than 1600 K, which is the result of the high luminosity of its host star. The TTV study of HAT-P-33b was analyzed by the transiting exoplanet monitoring project (TEMP) in the study of [44]. [44] refined HAT-P-33b orbital ephemerides and the system parameters through R-band photometric observations. Furthermore, the extended radial velocity measurements of this planet have been made by [45]. Their obtained transit parameters were consistent with the resulting parameters from [16], but smaller values for the ingress/egress duration, and the planet to star radius ratio ($R_P/R_*$) were deduced. These differences were because of derived complete light curves, which led to shorter and shallower transit shapes. However, they found no sign of anomalies in the TTV diagram. [46] performed additional follow-up ground-based photometric observations to confirm the HAT-P-33b planetary parameters through TEP modeling methods. [46] used a weighted linear least-squares analysis to update the reference ephemeris for this planet. Moreover, they found that the $R$-band transit depth in HAT-P-33b is larger than its discovery value [16] when discussing the variation in planetary radius against wavelength. [46]’s photometric data revealed no evidence of star-spot for this exoplanet.

We used the mid-transit times collected from the modeling of ETD and TESS light curves to prepare a new TTV diagram for HAT-P-33b. For this exoplanet, 30 mid-transit times from TESS sectors 20, 44, 45, 46, and 47 were extracted, also with seven mid-transit times from ETD observations.

\subsection{HAT-P-52b}
HAT-P-52b is a short-period gas-giant hot Jupiter discovered and characterized in 2015 by the transit method [18]. [47] reported new photometric light curves for HAT-P-52b and refined this system’s parameters by analyzing their light curves, previous photometric, and Doppler velocimetric data. [47] also performed the TTV study for HAT-P-52b and their results represented no significant trend in the TTV diagram and the measured mid-transit time was consistent with their updated linear ephemerides.

We plotted the updated TTV diagram based on the seven mid-transit times from the modeling of ETD and 10 mid-transit times from the TESS light curves (sectors 43 and 44), along with published data in previous literature.

\section{Summary and Conclusion}
We conducted a study on seven HATNet survey-selected transiting exoplanets and plotted the TTV diagrams. The goal of this study is to improve the planetary systems' reference ephemerides and to discuss the reasons for the period variations in these systems for future studies.

We have presented a new ephemeris for each of the seven exoplanets. For this purpose, we utilised the mid-transit times found in the literature as well as the light curves observed by ETD, TESS, and our ground-based observations. We used 11 mid-transit times from our observations in this study, which were made at eight observatories from 2018 to 2022.

Exofast-v1 was used to model the available light curves and extract the mid-transit times. We used the MCMC method to plot new TTV diagrams and refine exoplanets' ephemeris. The TTV diagrams show the orbital periods of exoplanets HAT-P-6b, HAT-P-18b, HAT-P-22b, HAT-P-32b, HAT-P-33b, and HAT-P-52b are increasing, whereas exoplanet HAT-P-12b has a declining tendency. It is probable that the six exoplanets' orbital periods increased since their ephemeris accuracy has become inaccurate over time.

According to the new ephemeris for exoplanet HAT-P-6b, it seems that the uncertainties of $t$ and $P$ should be more carefully considered in future investigations and observations.

\begin{table*}
\caption{Transit times of HAT-P-6b.}
\centering
\begin{center}
\footnotesize
\begin{tabular}{c c c c c c c c c c}
 \hline
 \hline
 $T_c(BJD_{TDB})$ & Error & Epoch & O-C & Ref./ETD Observer & $T_c(BJD_{TDB})$ & Error & Epoch & O-C & Ref./ETD Observer\\
\hline
2454035.67648	&	0.00027	&	0	&	0.0000	&	[19]	&	2455160.75292	&	0.00034	&	292	&	0.0048	&	KPNO 2.1m\\
2454035.67652	&	0.00028	&	0	&	0.0000	&	[14]	&	2455430.45630	&	0.00180	&	362	&	-0.0008	&	[20]\\
2454347.76839	&	0.00042	&	81	&	0.0001	&	[19]	&	2455430.46118	&	0.00100	&	362	&	0.0041	&	R. Dřevěný\\
2454698.39160	&	0.00110	&	172	&	0.0017	&	[19]	&	2455430.46570	&	0.00130	&	362	&	0.0086	&	TRESCA\\
2454740.77668	&	0.00063	&	183	&	0.0039	&	UDEM	&	2455430.46701	&	0.00100	&	362	&	0.0099	&	P. Veres\\
2455064.42616	&	0.00119	&	267	&	0.0026	&	L. Brát	&	2456193.35362	&	0.00077	&	560	&	0.0055	&	Poddaný, Moudrá\\
2455064.42751	&	0.00089	&	267	&	0.0040	&	J. Trnka, M. Klos	&	2456208.76626	&	0.00063	&	564	&	0.0062	&	J. Garlitz\\
2455064.42843	&	0.00061	&	267	&	0.0049	&	Brát et al.	&	2457603.55511	&	0.00111	&	926	&	0.0145	&	J.L. Salto\\
2455064.43067	&	0.00134	&	267	&	0.0072	&	R. Dřevěný, T. Kalisch	&	2459495.37153	&	0.00076	&	1417	&	0.0153	&	M. Raetz\\
\hline
\hline
\end{tabular}
\end{center}
\label{tab5}
\end{table*}

\begin{table*}
\caption{Transit times of HAT-P-18b.}
\centering
\begin{center}
\footnotesize
\begin{tabular}{c c c c c c c c c c}
 \hline
 \hline
 $T_c(BJD_{TDB})$ & Error & Epoch & O-C & Ref./ETD Observer & $T_c(BJD_{TDB})$ & Error & Epoch & O-C & Ref./ETD Observer\\
\hline
2454715.02174	&	0.00020	&	0	&	0.0000	&	[29]	& 2456455.56573	&	0.00075	&	316	&	0.0087	&	R. Naves	\\
2455662.40408	&	0.00054	&	172	&	0.0024	&	V.P. Hentunen	& 2456780.53283	&	0.00089	&	375	&	0.0025	&	F.G. Horta	\\
2455673.41967	&	0.00124	&	174	&	0.0019	&	[31]	& 2456780.53350	&	0.00073	&	375	&	0.0031	&	CAAT	\\
2455695.45273	&	0.00081	&	178	&	0.0029	&	A. Ayiomamitis	& 2456780.53981	&	0.00128	&	375	&	0.0094	&	A. Carreno	\\
2455706.46499	&	0.00077	&	180	&	-0.0009	&	C. Lopresti	& 2457116.52736	&	0.00073	&	436	&	0.0076	&	J. Gonzalez	\\
2455706.46993	&	0.00080	&	180	&	0.0040	&	[31]	& 2457149.56917	&	0.00092	&	442	&	0.0013	&	M. Deldem	\\
2455717.48233	&	0.00177	&	182	&	0.0004	&	A.L. Marrero	& 2457276.25646	&	0.00010	&	465	&	0.0040	&	[35]	\\
2455728.49780	&	0.00100	&	184	&	-0.0002	&	V. Benishek	& 2457474.54607	&	0.00046	&	501	&	0.0048	&	Signoret, Pioppa	\\
2455739.52081	&	0.00080	&	186	&	0.0068	&	F.C. Pecharromán	& 2457485.56256	&	0.00067	&	503	&	0.0053	&	Molina, Sureda	\\
2456042.45973	&	0.00124	&	241	&	0.0044	&	F.G. Horta	& 2457485.56424	&	0.00084	&	503	&	0.0069	&	D. Molina	\\
2456053.47276	&	0.00084	&	243	&	0.0014	&	[31]	& 2457507.59220	&	0.00019	&	507	&	0.0028	&	[32]	\\
2456086.51856	&	0.00125	&	249	&	-0.0009	&	[31]	& 2457843.58653	&	0.00090	&	568	&	0.0077	&	Scaggiante, Zardin	\\
2456097.53468	&	0.00074	&	251	&	-0.0008	&	T. Scarmato	& 2458559.62466	&	0.00111	&	698	&	0.0029	&	S. Dufoer	\\
2456400.48452	&	0.00108	&	306	&	0.0077	&	V. Popov	& 2458730.37292	&	0.00129	&	729	&	0.0024	&	B. Andreas	\\
2456411.49638	&	0.00084	&	308	&	0.0036	&	[31]	& 2459352.78320	&	0.00090	&	842	&	0.0061	&	[33]	\\
\hline
\hline
\end{tabular}
\end{center}
\label{tab6}
\end{table*}

\begin{table*}
\caption{Transit times of HAT-P-22b.}
\centering
\begin{center}
\footnotesize
\begin{tabular}{c c c c c c c c c c}
 \hline
 \hline
 $T_c(BJD_{TDB})$ & Error & Epoch & O-C & Ref./ETD Observer & $T_c(BJD_{TDB})$ & Error & Epoch & O-C & Ref./ETD Observer\\
\hline
2454930.22001	&	0.00025	&	0	&	0.0000	&	[17]	&	2457734.50182	&	0.00044	&	873	&	0.0137	&	V.P. Hentunen	\\
2455601.57849	&	0.00079	&	209	&	0.0045	&	L. Brát	&	2457779.47248	&	0.00075	&	887	&	0.0133	&	D. Molina	\\
2455614.43147	&	0.00126	&	213	&	0.0086	&	L. Brát	&	2457840.50512	&	0.00099	&	906	&	0.0138	&	M. Raetz	\\
2455694.73397	&	0.00112	&	238	&	0.0056	&	S. Shadic, M. Butler	&	2458171.36382	&	0.00094	&	1009	&	0.0138	&	V.P. Hentunen	\\
2455935.64851	&	0.00113	&	313	&	0.0036	&	S. Shadic	&	2458537.55631	&	0.00064	&	1123	&	0.0132	&	F. Campos	\\
2455948.50056	&	0.00126	&	317	&	0.0068	&	G. Marino	&	2458550.40688	&	0.00053	&	1127	&	0.0149	&	J. Gaitan	\\
2456298.63509	&	0.00098	&	426	&	0.0094	&	G.M. Schteinman	&	2458929.44872	&	0.00075	&	1245	&	0.0148	&	M. Theusner	\\
2456327.53684	&	0.00066	&	435	&	0.0011	&	J. Lozano de Haro	&	2458945.50651	&	0.00071	&	1250	&	0.0115	&	M. Theusner	\\
2456346.81517	&	0.00140	&	441	&	0.0061	&	[38]	&	2458945.51538	&	0.00117	&	1250	&	0.0204	&	A. Wunsche	\\
2456356.45198	&	0.00059	&	444	&	0.0063	&	A. Ayiomamitis	&	2458945.51541	&	0.00055	&	1250	&	0.0204	&	Y. Jongen	\\
2456391.78736	&	0.00088	&	455	&	0.0072	&	S. Shadic	&	2458945.51741	&	0.00106	&	1250	&	0.0224	&	M. Raetz	\\
2456671.25040	&	0.00020	&	542	&	0.0071	&	[36]	&	2458974.42843	&	0.00083	&	1259	&	0.0234	&	Scaggiante, Zardin	\\
2456687.31391	&	0.00055	&	547	&	0.0096	&	V.P. Hentunen	&	2459202.49043	&	0.00142	&	1330	&	0.0178	&	V. Perroud	\\
2456706.58936	&	0.00090	&	553	&	0.0117	&	[37]	&	2459276.36752	&	0.00079	&	1353	&	0.0138	&	Scaggiante, Zardin	\\
2456738.70941	&	0.00062	&	563	&	0.0095	&	[38]	&	2459276.37348	&	0.00044	&	1353	&	0.0198	&	M. Raetz	\\
2457024.58745	&	0.00090	&	652	&	0.0000	&	CAAT	&	2459308.49480	&	0.00049	&	1363	&	0.0189	&	M. Bachschmidt	\\
2457024.59630	&	0.00105	&	652	&	0.0089	&	R.N. Nogues	&	2459308.49839	&	0.00110	&	1363	&	0.0225	&	Y. Jongen	\\
2457127.38972	&	0.00132	&	684	&	0.0112	&	A. Valdera	& \\
\hline
\hline
\end{tabular}
\end{center}
\label{tab7}
\end{table*}

\begin{table*}
\caption{Transit times of HAT-P-12b.}
\centering
\begin{center}
\footnotesize
\begin{tabular}{c c c c c c c c c c}
 \hline
 \hline
 $T_c(BJD_{TDB})$ & Error & Epoch & O-C & Ref./ETD Observer &  $T_c(BJD_{TDB})$ & Error & Epoch & O-C & Ref./ETD Observer\\
\hline
2454187.85647	&	0.00061	&	-72	&	0.0012	&	[15] &	2456732.59827	&	0.00063	&	720	&	-0.0003	&	[24]	\\
2454216.77343	&	0.00023	&	-63	&	0.0006	&	[15] &	2456735.81023	&	0.00029	&	721	&	-0.0014	&	[27]	\\
2454419.19556	&	0.00020	&	0	&	0.0000	&	[15]	&	2456745.45299	&	0.00046	&	724	&	0.0021	&	G. Corfini	\\
2454869.02462	&	0.00033	&	140	&	0.0007	&	[15]	&	2456745.45375	&	0.00074	&	724	&	0.0029	&	A. Valdera	\\
2454897.94297	&	0.00057	&	149	&	0.0015	&	[15]	&	2456764.72820	&	0.00040	&	730	&	-0.0010	&	[27]	\\
2454965.41502	&	0.00062	&	170	&	-0.0007	&	H. Kučáková	&	2456793.64833	&	0.00092	&	739	&	0.0016	&	[24]	\\
2455347.76929	&	0.00021	&	289	&	-0.0006	&	[21]	&	2456806.49642	&	0.00062	&	743	&	-0.0026	&	[24]	\\
2455630.51929	&	0.00011	&	377	&	0.0002	&	[24]	&	2456806.49815	&	0.00019	&	743	&	-0.0008	&	[24]	\\
2455646.58184	&	0.00049	&	382	&	-0.0026	&	Š. Gajdoš, J. Világi	&	2456809.71238	&	0.00045	&	744	&	0.0003	&	[23]	\\
2455649.79746	&	0.00020	&	383	&	0.0000	&	[27]	&	2457063.54386	&	0.00040	&	823	&	0.0001	&	M. Bretton	\\
2455649.79770	&	0.00035	&	383	&	0.0002	&	[22]	&	2457124.59055	&	0.00039	&	842	&	-0.0014	&	M. Bretton	\\
2455659.43649	&	0.00044	&	386	&	-0.0002	&	[24]	&	2457217.76896	&	0.00057	&	871	&	-0.0017	&	[27]	\\
2455659.43650	&	0.00037	&	386	&	-0.0001	&	J. Ruiz	&	2457516.58394	&	0.00031	&	964	&	-0.0013	&	[25]	\\
2455665.86227	&	0.00037	&	388	&	-0.0005	&	[22]	&	2457529.43692	&	0.00016	&	968	&	-0.0005	&	M. Bretton	\\
2455669.07486	&	0.00080	&	389	&	-0.0010	&	[27]	&	2457831.46339	&	0.00037	&	1062	&	-0.0017	&	[25]	\\
2455675.50436	&	0.00047	&	391	&	0.0024	&	[24]	&	2457834.67573	&	0.00042	&	1063	&	-0.0024	&	[25]	\\
2455678.71382	&	0.00041	&	392	&	-0.0012	&	[27]	&	2457837.88977	&	0.00095	&	1064	&	-0.0014	&	[26]	\\
2455678.71462	&	0.00047	&	392	&	-0.0004	&	[23]	&	2457847.53037	&	0.00025	&	1067	&	0.0000	&	M. Bretton	\\
2455691.56661	&	0.00022	&	396	&	-0.0006	&	[24]	&	2457876.44558	&	0.00036	&	1076	&	-0.0023	&	[25]	\\
2455694.78089	&	0.00029	&	397	&	0.0006	&	[22]	&	2457879.66160	&	0.00036	&	1077	&	0.0006	&	[25]	\\
2455707.63244	&	0.00027	&	401	&	-0.0001	&	[24]	&	2457892.51113	&	0.00046	&	1081	&	-0.0021	&	[25]	\\
2455720.48443	&	0.00015	&	405	&	-0.0003	&	[24]	&	2457892.51191	&	0.00012	&	1081	&	-0.0013	&	[25]	\\
2455736.54999	&	0.00030	&	410	&	-0.0001	&	[24]	&	2457892.51258	&	0.00041	&	1081	&	-0.0006	&	[25]	\\
2455964.67787	&	0.00059	&	481	&	0.0005	&	[24]	&	2457908.57706	&	0.00040	&	1086	&	-0.0014	&	[25]	\\
2456006.44644	&	0.00062	&	494	&	-0.0007	&	V. Přibík	&	2457940.70759	&	0.00036	&	1096	&	-0.0015	&	[25]	\\
2456006.44779	&	0.00012	&	494	&	0.0007	&	[24]	&	2458223.46181	&	0.00092	&	1184	&	0.0034	&	V. Hentunen	\\
2456041.79018	&	0.00037	&	505	&	-0.0006	&	[23]	&	2458541.55104	&	0.00042	&	1283	&	-0.0002	&	Y. Jongen	\\
2456051.42937	&	0.00048	&	508	&	-0.0006	&	[24]	&	2458570.46838	&	0.00042	&	1292	&	-0.0004	&	Y. Jongen	\\
2456054.64202	&	0.00060	&	509	&	-0.0010	&	[24]	&	2458859.64077	&	0.00055	&	1382	&	-0.0034	&	V. Hentunen	\\
2456070.70792	&	0.00035	&	514	&	-0.0004	&	[23]	&	2458904.62566	&	0.00033	&	1396	&	-0.0014	&	Y. Jongen	\\
2456083.55936	&	0.00065	&	518	&	-0.0012	&	[24]	&	2458933.54260	&	0.00047	&	1405	&	-0.0020	&	M. Raetz	\\
2456086.77329	&	0.00051	&	519	&	-0.0003	&	[23]	&	2458933.54500	&	0.00064	&	1405	&	0.0004	&	Y. Jongen	\\
2456359.88215	&	0.00014	&	604	&	-0.0015	&	[28]	&	2458933.54535	&	0.00076	&	1405	&	0.0008	&	R.F. Auer	\\
2456363.09540	&	0.00019	&	605	&	-0.0013	&	[28]	&	2458946.39354	&	0.00065	&	1409	&	-0.0033	&	M. Raetz	\\
2456388.80064	&	0.00048	&	613	&	-0.0006	&	[24]	&	2458946.39547	&	0.00062	&	1409	&	-0.0013	&	Y. Jongen	\\
2456388.80101	&	0.00081	&	613	&	-0.0002	&	[24]	&	2458946.39639	&	0.00043	&	1409	&	-0.0004	&	M. Bretton	\\
2456462.70081	&	0.00050	&	636	&	-0.0008	&	[23]	&	2458946.39685	&	0.00104	&	1409	&	0.0000	&	F. Salvaggio	\\
2456716.53165	&	0.00033	&	715	&	-0.0017	&	[24]	&	2458962.46093	&	0.00031	&	1414	&	-0.0012	&	M. Raetz	\\
2456732.59741	&	0.00023	&	720	&	-0.0012	&	[24]	&	2459296.61925	&	0.00038	&	1518	&	-0.0011	&	Y. Jongen	\\
2456732.59765	&	0.00011	&	720	&	-0.0010	&	[24]	&	2459309.47053	&	0.00046	&	1522	&	-0.0020	&	M. Bachschmidt	\\
2456732.59796	&	0.00009	&	720	&	-0.0007	&	[24]	&	2459309.47209	&	0.00044	&	1522	&	-0.0005	&	Montigiani, Mannucci	\\
2456732.59811	&	0.00028	&	720	&	-0.0005	&	[24]	&	2459354.45314	&	0.00034	&	1536	&	-0.0023	&	Y. Jongen	\\
\hline
\hline
\end{tabular}
\end{center}
\label{tab8}
\end{table*}

\begin{table*}
\caption{Transit times of HAT-P-52b.}
\centering
\begin{center}
\footnotesize
\begin{tabular}{c c c c c c c c c c}
 \hline
 \hline
 $T_c(BJD_{TDB})$ & Error & Epoch & O-C & Ref./ETD Observer & $T_c(BJD_{TDB})$ & Error & Epoch & O-C & Ref./ETD Observer\\
\hline
2455852.10326	&	0.00041	&	0	&	0.0000	&	[18]	&	2457727.30532	&	0.00049	&	681	&	0.0037	&	M. Bretton	\\
2455995.29102	&	0.00031	&	52	&	0.0008	&	[47]	&	2457749.32972	&	0.00193	&	689	&	-0.0007	&	[48]	\\
2457311.51042	&	0.00142	&	530	&	0.0017	&	[48]	&	2458060.48649	&	0.00155	&	802	&	-0.0002	&	[48]	\\
2457336.29302	&	0.00277	&	539	&	0.0019	&	[48]	&	2458140.34268	&	0.00137	&	831	&	0.0017	&	[48]	\\
2457336.29562	&	0.00164	&	539	&	0.0045	&	M. Bretton	&	2458140.34636	&	0.00214	&	831	&	0.0054	&	[48]	\\
2457399.62421	&	0.00133	&	562	&	0.0004	&	[48]	&	2458429.47243	&	0.00066	&	936	&	0.0040	&	P. Guerra	\\
2457413.39279	&	0.00091	&	567	&	0.0010	&	[48]	&	2458531.35748	&	0.00170	&	973	&	0.0060	&	M. Raetz	\\
2457724.55292	&	0.00121	&	680	&	0.0049	&	R. Naves	&	2459167.44167	&	0.00098	&	1204	&	0.0097	&	S. Gudmundsson	\\
2457727.30439	&	0.00107	&	681	&	0.0027	&	[48]	&	2459514.39070	&	0.00048	&	1330	&	0.0057	&	A. Wünsche	\\
\hline
\hline
\end{tabular}
\end{center}
\label{tab9}
\end{table*}

\begin{table*}
\caption{Transit times of HAT-P-32b.}
\centering
\begin{center}
\footnotesize
\begin{tabular}{c c c c c c c c c c}
 \hline
 \hline
 $T_c(BJD_{TDB})$ & Error & Epoch & O-C & Ref./ETD Observer &  $T_c(BJD_{TDB})$ & Error & Epoch & O-C & Ref./ETD Observer\\
\hline
2454420.44637	&	0.00009	&	0	&	0.0000	&	[16]	&	2456600.55546	&	0.00017	&	1014	&	0.0010	&	[41]	\\
2455796.45134	&	0.00092	&	640	&	-0.0001	&	Moudrá, Sobotka	&	2456600.55612	&	0.00033	&	1014	&	0.0016	&	P. Benni	\\
2455839.45347	&	0.00101	&	660	&	0.0018	&	[41]	&	2456611.30423	&	0.00064	&	1019	&	-0.0003	&	F.G. Horta	\\
2455843.75341	&	0.00019	&	662	&	0.0017	&	[41]	&	2456615.60935	&	0.00034	&	1021	&	0.0048	&	M. Perchak	\\
2455843.75341	&	0.00019	&	662	&	0.0017	&	[21]	&	2456617.75617	&	0.00042	&	1022	&	0.0016	&	S. Shadick	\\
2455845.90287	&	0.00024	&	663	&	0.0012	&	[41]	&	2456628.50585	&	0.00031	&	1027	&	0.0013	&	[41]	\\
2455845.90287	&	0.00024	&	663	&	0.0012	&	[21]	&	2456637.10480	&	0.00023	&	1031	&	0.0002	&	[42]	\\
2455845.90314	&	0.00040	&	663	&	0.0015	&	[41]	&	2456639.25687	&	0.00046	&	1032	&	0.0022	&	Aldi, Leo, Bozza	\\
2455845.90314	&	0.00040	&	663	&	0.0015	&	[21]	&	2456656.45533	&	0.00045	&	1040	&	0.0006	&	[41]	\\
2455852.35180	&	0.00073	&	666	&	0.0001	&	A. Brosio	&	2456656.45798	&	0.00064	&	1040	&	0.0033	&	R. Naves	\\
2455856.65098	&	0.00029	&	668	&	-0.0007	&	S. Shadick	&	2456957.45714	&	0.00044	&	1180	&	0.0013	&	E.D. Alonso	\\
2455858.80105	&	0.00039	&	669	&	-0.0007	&	S. Shadick	&	2456959.60773	&	0.00024	&	1181	&	0.0019	&	C.G.J. Dibon	\\
2455860.95425	&	0.00034	&	670	&	0.0025	&	S. Shadick	&	2456972.50716	&	0.00069	&	1187	&	0.0013	&	F.G. Horta	\\
2455867.40301	&	0.00073	&	673	&	0.0013	&	[41]	&	2456998.30682	&	0.00029	&	1199	&	0.0009	&	M. Salisbury	\\
2455880.30170	&	0.00035	&	679	&	-0.0001	&	G. Corfini	&	2457009.05775	&	0.00080	&	1204	&	0.0017	&	[42]	\\
2455880.30267	&	0.00033	&	679	&	0.0009	&	[41]	&	2457013.35882	&	0.00042	&	1206	&	0.0028	&	F. Campos	\\
2455895.35249	&	0.00080	&	686	&	0.0006	&	[41]	&	2457024.10743	&	0.00046	&	1211	&	0.0014	&	[42]	\\
2455895.35297	&	0.00016	&	686	&	0.0011	&	[41]	&	2457245.55959	&	0.00028	&	1314	&	0.0027	&	M. Bretton	\\
2455897.50328	&	0.00033	&	687	&	0.0014	&	[41]	&	2457286.40911	&	0.00032	&	1333	&	0.0021	&	M. Salisbury	\\
2455910.40274	&	0.00043	&	693	&	0.0008	&	[41]	&	2457303.60475	&	0.00055	&	1341	&	-0.0024	&	F. Signoret	\\
2455912.54988	&	0.00147	&	694	&	-0.0020	&	R. Naves	&	2457342.30813	&	0.00054	&	1359	&	0.0009	&	M. Bretton	\\
2455923.30295	&	0.00031	&	699	&	0.0010	&	[41]	&	2457372.41074	&	0.00069	&	1373	&	0.0034	&	D. Molina	\\
2455940.50354	&	0.00109	&	707	&	0.0015	&	R. Naves	&	2457385.31188	&	0.00036	&	1379	&	0.0045	&	M. Bretton	\\
2455942.65179	&	0.00113	&	708	&	-0.0002	&	[41]	&	2457396.05803	&	0.00091	&	1384	&	0.0006	&	[42]	\\
2455942.65287	&	0.00064	&	708	&	0.0008	&	[41]	&	2457415.40279	&	0.00081	&	1393	&	-0.0047	&	J. Gaitan	\\
2455942.65443	&	0.00051	&	708	&	0.0024	&	D. Zaharevitz	&	2457660.50761	&	0.00057	&	1507	&	-0.0008	&	R. Roy	\\
2456155.50385	&	0.00026	&	807	&	0.0010	&	[41]	&	2457660.50931	&	0.00049	&	1507	&	0.0009	&	M. Deldem	\\
2456157.65470	&	0.00072	&	808	&	0.0019	&	[41]	&	2457664.80650	&	0.00255	&	1509	&	-0.0019	&	M. Fowler	\\
2456170.55416	&	0.00089	&	814	&	0.0013	&	F. Centenera	&	2457692.76393	&	0.00187	&	1522	&	0.0054	&	M. Fowler	\\
2456177.00392	&	0.00025	&	817	&	0.0010	&	[40]	&	2457725.00958	&	0.00017	&	1537	&	0.0009	&	W. Kang	\\
2456183.45361	&	0.00049	&	820	&	0.0007	&	[41]	&	2457759.41008	&	0.00034	&	1553	&	0.0013	&	V.P. Hentunen	\\
2456183.45364	&	0.00085	&	820	&	0.0007	&	[41]	&	2457772.30878	&	0.00034	&	1559	&	-0.0001	&	M. Bretton	\\
2456185.60375	&	0.00033	&	821	&	0.0008	&	[41]	&	2458004.51187	&	0.00057	&	1667	&	0.0022	&	F.G. Horta	\\
2456211.40267	&	0.00214	&	833	&	-0.0004	&	[41]	&	2458021.71057	&	0.00027	&	1675	&	0.0008	&	O. Cooper	\\
2456211.40361	&	0.00056	&	833	&	0.0006	&	[41]	&	2458088.36081	&	0.00032	&	1706	&	0.0008	&	P. Guerra	\\
2456220.00440	&	0.00019	&	837	&	0.0013	&	[40]	&	2458107.71300	&	0.00071	&	1715	&	0.0029	&	[43]	\\
2456226.45618	&	0.00102	&	840	&	0.0031	&	[41]	&	2458116.31178	&	0.00030	&	1719	&	0.0017	&	F. Bruno	\\
2456228.60389	&	0.00057	&	841	&	0.0008	&	A. Carreño	&	2458367.86227	&	0.00074	&	1836	&	0.0012	&	B. Rodgers	\\
2456232.90554	&	0.00028	&	843	&	0.0024	&	Kehusmaa, Harlingten	&	2458376.46386	&	0.00042	&	1840	&	0.0028	&	G. Rousseau	\\
2456237.20386	&	0.00029	&	845	&	0.0007	&	[42]	&	2458391.51025	&	0.00020	&	1847	&	-0.0009	&	M. Bretton	\\
2456239.35339	&	0.00050	&	846	&	0.0002	&	J. Gaitan	&	2458404.40968	&	0.00081	&	1853	&	-0.0015	&	A. Wünsche	\\
2456239.35397	&	0.00052	&	846	&	0.0008	&	F.G. Horta	&	2458406.56109	&	0.00049	&	1854	&	-0.0001	&	V.P. Hentunen	\\
2456243.65354	&	0.00086	&	848	&	0.0004	&	R. Naves	&	2458447.41287	&	0.00067	&	1873	&	0.0015	&	Y. Jongen	\\
2456250.10439	&	0.00042	&	851	&	0.0012	&	[42]	&	2458490.41085	&	0.00040	&	1893	&	-0.0007	&	S.F. Dagot	\\
2456254.40251	&	0.00063	&	853	&	-0.0007	&	R. Naves	&	2458503.31396	&	0.00031	&	1899	&	0.0024	&	Y. Jongen	\\
2456254.40404	&	0.00022	&	853	&	0.0008	&	[41]	&	2458518.36601	&	0.00067	&	1906	&	0.0044	&	D. Laloum	\\
2456265.15466	&	0.00048	&	858	&	0.0014	&	[42]	&	2458720.46171	&	0.00129	&	2000	&	-0.0007	&	V. Perroud	\\
2456284.50460	&	0.00100	&	867	&	0.0013	&	[41]	&	2458739.81430	&	0.00382	&	2009	&	0.0019	&	[43]	\\
2456297.39940	&	0.00069	&	873	&	-0.0040	&	R. Naves	&	2458821.51375	&	0.00044	&	2047	&	0.0010	&	Y. Jongen	\\
2456514.55505	&	0.00065	&	974	&	0.0009	&	F.G. Horta	&	2458847.31493	&	0.00044	&	2059	&	0.0021	&	J. Dirscherl	\\
2456527.45567	&	0.00030	&	980	&	0.0015	&	V.P. Hentunen	&	2458849.46424	&	0.00031	&	2060	&	0.0014	&	Y. Jongen	\\
2456542.50522	&	0.00052	&	987	&	0.0010	&	[41]	&	2458862.36411	&	0.00035	&	2066	&	0.0012	&	Y. Jongen	\\
2456542.50530	&	0.00018	&	987	&	0.0010	&	[41]	&	2459109.61612	&	0.00038	&	2181	&	0.0023	&	Y. Jongen	\\
2456542.50538	&	0.00032	&	987	&	0.0011	&	[41]	&	2459120.36332	&	0.00054	&	2186	&	-0.0005	&	V.P. Hentunen	\\
2456557.55814	&	0.00057	&	994	&	0.0038	&	Herrero, Pascual	&	2459178.41494	&	0.00085	&	2213	&	0.0009	&	F. Mortari	\\
2456559.70548	&	0.00025	&	995	&	0.0011	&	P. Benni	&	2459197.76032	&	0.00152	&	2222	&	-0.0038	&	R. Dymock	\\
2456572.60532	&	0.00018	&	1001	&	0.0009	&	[41]	&	2459221.41318	&	0.00066	&	2233	&	-0.0011	&	S. Gudmundsson	\\
2456585.50553	&	0.00122	&	1007	&	0.0011	&	J. Mravik, J. Grnja	&	2459481.56450	&	0.00054	&	2354	&	-0.0007	&	Y. Jongen	\\
2456598.40539	&	0.00017	&	1013	&	0.0009	&	[41]	&	2459535.31403	&	0.00113	&	2379	&	-0.0014	&	D. Daniel	\\
\hline
\hline
\end{tabular}
\end{center}
\label{tab10}
\end{table*}

\begin{table*}
\caption{Transit times of HAT-P-33b.}
\centering
\begin{center}
\footnotesize
\begin{tabular}{c c c c c c c c c c}
 \hline
 \hline
 $T_c(BJD_{TDB})$ & Error & Epoch & O-C & Ref./ETD Observer & $T_c(BJD_{TDB})$ & Error & Epoch & O-C & Ref./ETD Observer\\
\hline
2454429.93117	&	0.00055	&	-196	&	0.0021	&	[16]	&	2456723.08352	&	0.00047	&	464	&	0.0016	&	[44]	\\
2454551.53949	&	0.00058	&	-161	&	0.0039	&	[16]	&	2456723.08417	&	0.00059	&	464	&	0.0023	&	[44]	\\
2454794.75180	&	0.00037	&	-91	&	0.0030	&	[16]	&	2457039.26162	&	0.00046	&	555	&	0.0026	&	[44]	\\
2454801.69988	&	0.00074	&	-89	&	0.0021	&	[16]	&	2457046.20981	&	0.00046	&	557	&	0.0018	&	[44]	\\
2455110.92595	&	0.00022	&	0	&	0.0000	&	[16]	&	2457053.15793	&	0.00047	&	559	&	0.0010	&	[44]	\\
2455607.77606	&	0.00060	&	143	&	0.0003	&	[16]	&	2457067.05749	&	0.00043	&	563	&	0.0027	&	[44]	\\
2455614.72513	&	0.00036	&	145	&	0.0005	&	[16]	&	2457397.13183	&	0.00066	&	658	&	0.0020	&	[44]	\\
2455857.93657	&	0.00067	&	215	&	-0.0013	&	S. Shadick	&	2458498.54660	&	0.00113	&	975	&	0.0085	&	P. Guerra	\\
2455864.88110	&	0.00078	&	217	&	-0.0057	&	S. Shadick	&	2459172.58691	&	0.00093	&	1169	&	0.0009	&	V. Ferrando	\\
2456024.71746	&	0.00120	&	263	&	0.0048	&	[46]	&	2459172.58950	&	0.00078	&	1169	&	0.0034	&	Y. Jongen	\\
2456636.22181	&	0.00072	&	439	&	0.0018	&	[44]	&	2459172.59843	&	0.00148	&	1169	&	0.0124	&	A. Wünsche	\\
2456716.13465	&	0.00069	&	462	&	0.0017	&	[44]	&	2459266.39915	&	0.00068	&	1196	&	0.0023	&	M. Raetz	\\
2456716.13498	&	0.00065	&	462	&	0.0020	&	[44]	&	\\
\hline
\hline
\end{tabular}
\end{center}
\label{tab11}
\end{table*}

\section*{Acknowledgements}
We made use of the ETD data for this investigation. ETD is maintained by the Czech Astronomical Society (CAS), an old astronomical organization in the Czech Republic with more than hundreds of members (amateur and professional astronomers included).\\
This work includes data from the TESS mission observations. Funding for the TESS mission is provided by the NASA Explorer Program.

\section*{Data Availability}
The ground-based and TESS data used in this study are available upon request to the corresponding author.

\section*{ORCID iDs}
\noindent Atila Poro: 0000-0002-0196-9732\\
Farzaneh Ahangarani Farahani: 0000-0003-3740-7309\\
Esfandiar Jahangiri: 0000-0002-1576-798X\\
Ahmad Sarostad: 0000-0001-6485-8696\\
Mohammadjavad Gozarandi: 0000-0002-5131-6801\\
Marzie Haghgou: 0000-0002-1182-5871\\
Faezeh Abolhassani: 0000-0002-1235-4602\\
Amirhossein Fakhrabadi: 0000-0002-2735-7021\\
Anaël Wünsche: 0000-0002-6176-9847\\
Ramon Naves: 0000-0002-9463-9029\\
Pere Guerra: 0000-0002-4308-2339\\
Alessandro Marchini: 0000-0003-3779-6762\\
Mark Salisbury: 0009-0000-6809-7679\\
Roman Ehrenberger: 0000-0002-9720-9868\\

\vspace{1.5cm}

%%%%%%%%%%%%%%%%%%%% REFERENCES %%%%%%%%%%%%%%%%%%

%%%%%%%%%%%%%%%%%%%%%%%%%%%%%%%%%%%%%%%%%%%%%%%%%%

\bsp	% typesetting comment
\label{lastpage}
\end{document}